\documentclass[preprint,aps]{revtex4}
\usepackage{graphicx}
\usepackage{dcolumn}
\usepackage{bm}
\usepackage{color}
\usepackage{lipsum}
\begin{document}
\preprint{APS/123-QED}
\title{
Pressure induced evolution of band structure in black phosphorus studied by $^{31}$P-NMR
}
\author{T. Fujii}
\author{Y. Nakai}
\author{Y. Akahama}
\author{K. Ueda}
\author{T.~Mito}
 \email{mito@sci.u-hyogo.ac.jp}
\affiliation{
Graduate School of Material Science, University of Hyogo, Ako-gun 678-1297, Japan
}
\date{\today}

\begin{abstract}
Two-dimensional layered semiconductor black phosphorus (BP), a promising pressure induced Dirac system as predicted by band structure calculations, has been studied by $^{31}$P-nuclear magnetic resonance.
Band calculations have been also carried out to estimate the density of states $D(E)$.
The temperature and pressure dependences of nuclear spin lattice relaxation rate $1/T_1$ in the semiconducting phase are well reproduced using the derived $D(E)$, and the resultant pressure dependence of semiconducting gap is in good accordance with previous reports, giving a good confirmation that the band calculation on BP is fairly reliable.
The present analysis of $1/T_1$ data with the complemental theoretical calculations allows us to extract essential information, such as the pressure dependences of $D(E)$ and chemical potential, as well as to decompose observed $1/T_1$ into intrinsic and extrinsic contributions.
An abrupt increase in $1/T_1$ at 1.63~GPa indicates that the semiconducting gap closes, resulting in an enhancement of conductivity.
\end{abstract}

\pacs{71.27.+a, 75.20.Hr, 75.30.Mb, 78.70.Dm}
\maketitle


Novel electronic properties originating from conical band dispersion near the Fermi level, which is realized in Dirac and Wyle semimetals, have been recently intensively studied.
Among a growing number of candidate materials, the two-dimensional layered semiconductor black phosphorus (BP) presents ideal conditions to investigate systematically Dirac corn's formation from finite- to zero-gap states.
Transport and optical measurements indicate that the narrow band gap of approximately 0.3~eV is easily reduced to zero by applying pressure of $P_c = 1.2 \sim 1.5$~GPa \cite{Okajima,Akahama2,Xiang,Akiba}.
The appearance of small Fermi surfaces above $P_c$ is also evidenced by the observation of Shubnikov de Haas oscillation \cite{Xiang,Akiba}.
On the theoretical side, band calculations predict that four twofold-degenerate Dirac cores appear at the Z point at $P_c$ and BP becomes a 3D Dirac semimetal under hydrostatic pressure \cite{Gong}.

In this letter, we report the results of $^{31}$P-nuclear magnetic resonance (NMR) measurements on BP under pressure for the first time.
To investigate the band structure experimentally, the angle-resolved photoemission spectroscopy (ARPES) may be known as a powerful tool.
Indeed, ARPES measurements on BP consisting of a few stacking layers doped with potassium indicate the realization of Dirac semimetal with anisotropic dispersion \cite{Kim}.
However, in the case of undoped BP, making BP in the shape of a few-layer sample enhances the gap:
it moves away from the zero-gap state.
Thus the pressure is currently the most realistic external parameter to tune the band structure of BP toward the formation of Dirac cores without introducing additional impurities, and therefore one needs an experimental way to gain insight into the band structure even at high pressures.

Nuclear spin lattice relaxation rate $1/T_1$ measured by NMR reflects the square of the density of states, $D(E)^2$, for nonmagnetic materials.
In the previous studies of Dirac and Weyl semimetals such as $\alpha$-(BEDT-TTF)$_2$I$_3$ \cite{Hirata} and TaP \cite{Yasuoka}, respectively, the $T_1$ measurement was actually adopted to examine the existence of Dirac or Weyl nodes.
For the elemental semiconductor BP, due to its simple constituent, one can obtain the reliable estimation of $D(E)$ by conventional band calculations.
The analysis of $1/T_1$ data with the derived $D(E)$ model allows us to extract detailed information on the band structure that is not accessible without it.
We also note that, from the technical point of view, $P_c$ of BP is easily achievable for the NMR measurements.

\begin{figure}[t]
\includegraphics[width=0.6\linewidth]{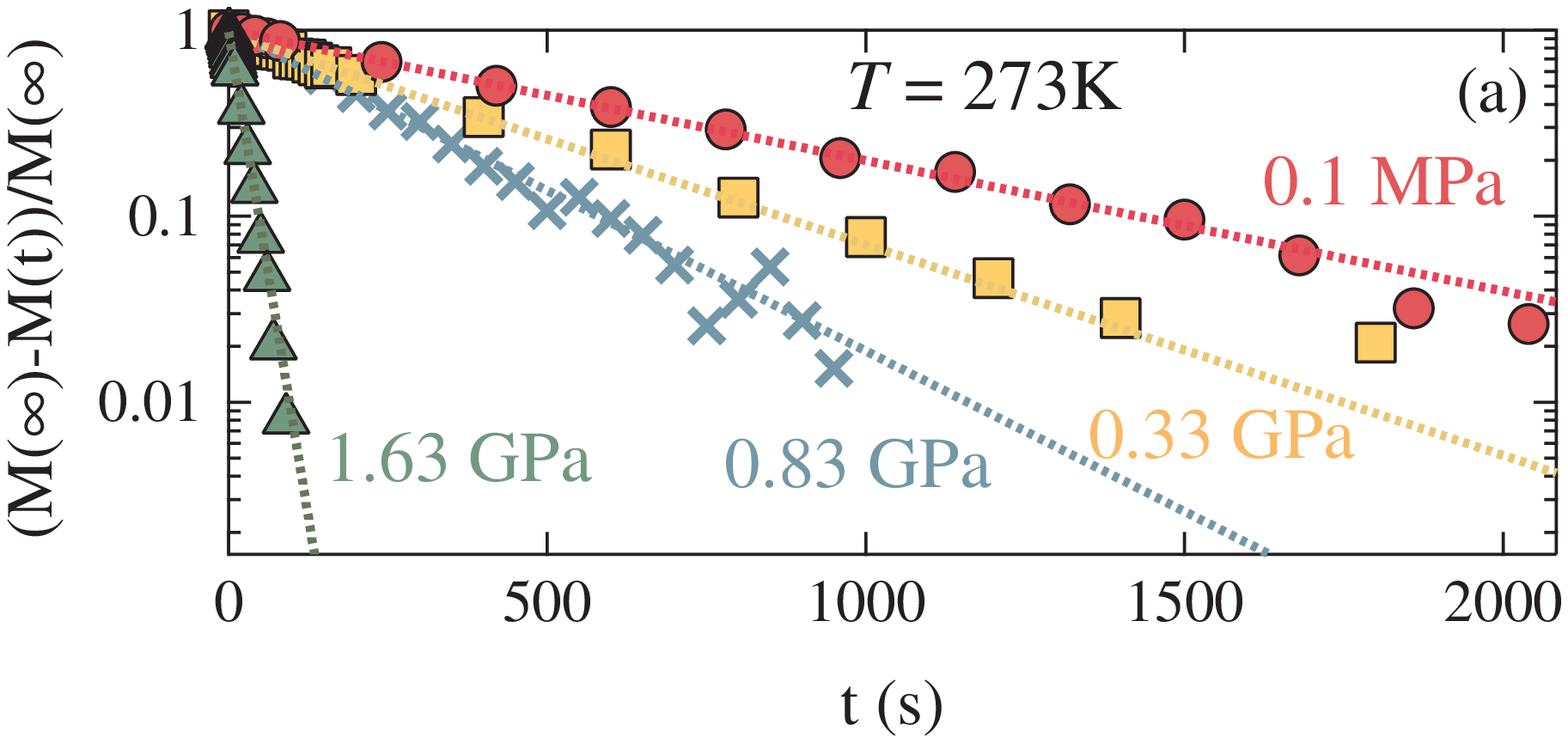}
\includegraphics[width=0.6\linewidth]{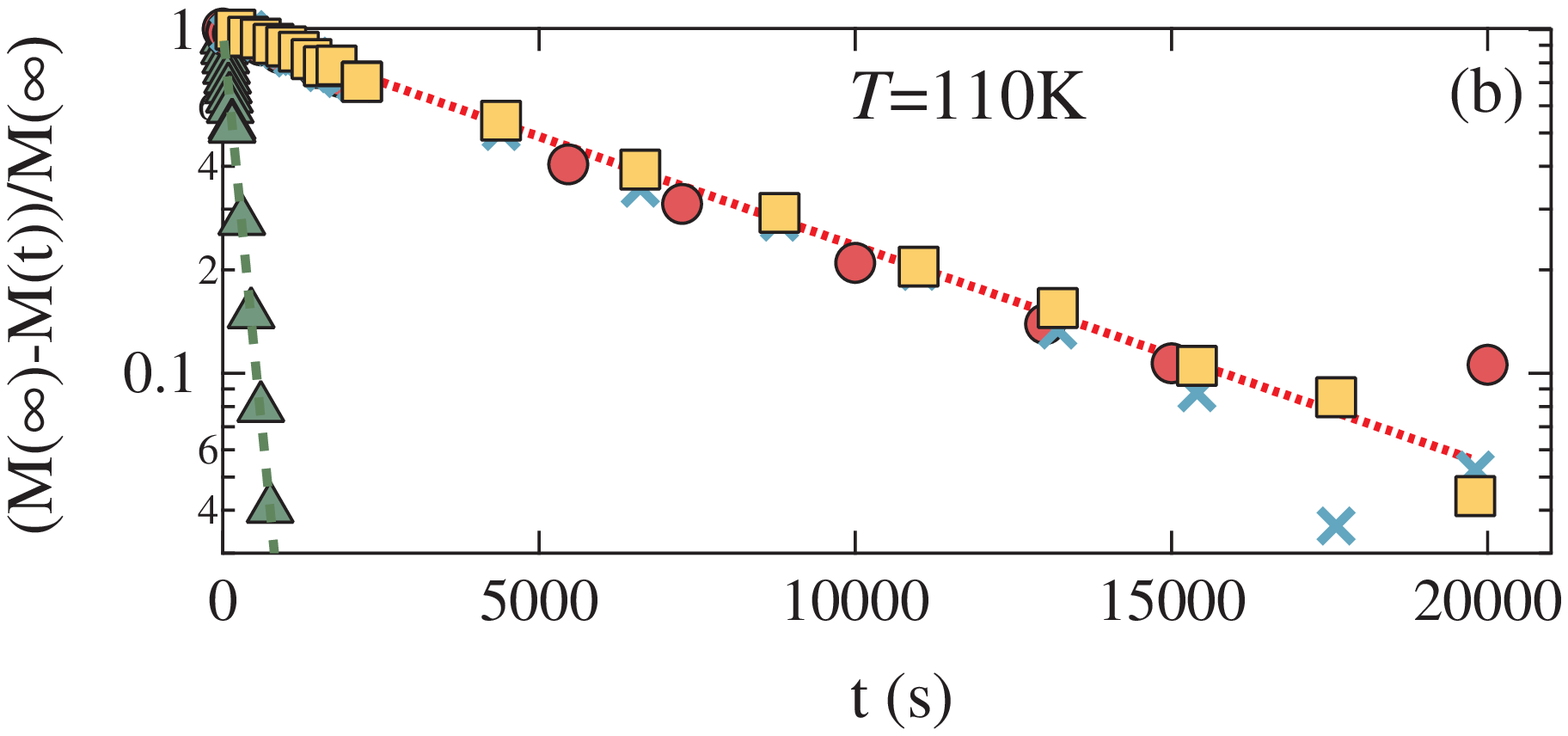}
\caption{\label{fig:epsart}
(color on line)
Pressure variation of $T_1$ relaxation curve $(M(\infty) - M(t))/ M(\infty)$ as a function of $t$ of BP.
(a) and (b) show the data at 273 and 110~K, respectively.
The solid lines are fits to a single exponential function.
}
\end{figure}

A polycrystalline sample of BP was prepared by a high-pressure synthesis technique \cite{Endo}. The high-pressure measurements of NMR were carried out using a self-clamped BeCu/NiCrAl piston-cylinder cell.
Daphne 7373 and silicon-based organic liquid were used as pressure media for the measurements below and above 1GPa, respectively.
The applied pressure was monitored by measuring the resistance of a manganin wire gauge and the superconducting temperature of a tin manometer.
The $^{31}$P-NMR spectra and $T_1$ were acquired by measuring the intensity of free induction decay signal with a phase-coherent pulsed spectrometer.
Band structure calculations were performed using WIEN2k package with the LDA + mBJ potential.
A $k$-mesh of $19 \times 19 \times 19$ was adopted to sample the first Brillouin zone.

\begin{figure}[t]
\includegraphics[width=0.6\linewidth]{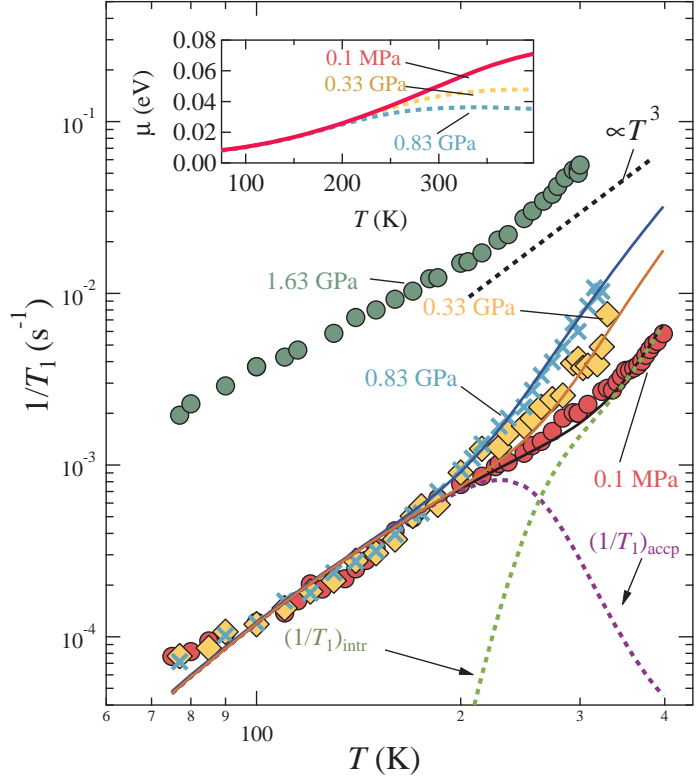}
\caption{\label{fig:epsart}
(color on line)
Temperature dependence of $1/T_1$ measured at ambient pressure, 0.33, 0.83, and 1.63 GPa.
The solid lines are fits the data to Eqs.~(1) and (2).
The broken lines are $(1/T_1)_{\rm intr}$ and $(1/T_1)_{\rm accp}$, which are decomposed from the data at ambient pressure.
See text for details.
Inset: temperature dependence of $\mu$ estimated for ambient pressure, 0.33~GPa, and 0.83~GPa.
}
\end{figure}

Obtained $^{31}$P-NMR spectral shape of BP is consistent with a previous report \cite{Bence}.
Figure~1 shows representative $T_1$ relaxation curves, where $M(t)$ is the nuclear magnetization at a delay time $t$ after saturation pulses.
The long decay time, for which $T_1$ reaches even $10^4$ sec below 100~K at ambient pressure (0.1~MPa) as shown later, leads to a difficulty in evaluating the accurate value of $T_1$.
For the reliable estimation of $T_1$, $M(t)$ was measured up to sufficiently long $t$'s, and all the relaxation curves in the present experiment are found to follow a single exponential function as expected for nuclear spin $1/2$.
As indicated in Fig.~1(a), $T_1$ at 273~K is monotonically shortened with increasing pressure.
In contrast, Fig.~1(b) reveals that the relaxation curve at 110~K hardly depends on pressure up to 0.83~GPa.

The obtained temperature dependence of $1/T_1$ at different pressures is shown in Fig.~2.
In the pressure range up to 0.83~GPa, where BP is in the semiconducting phase, $1/T_1$ above 200~K increases with pressure, while $1/T_1$ below 200~K is almost independent of pressure.
These characteristics are exactly expected from the results of Figs.~1(a) and (b).
By considering previous reports that 
(i) the resistivity and the Hall mobility show intrinsic behavior of usual semiconductors above 350~K \cite{Akahama1},
(ii) the semiconducting gap is reduced with pressure \cite{Akahama2,Xiang,Akiba,Chun-Hong}, and
(iii) the undoped BP sample is a $p$-type semiconductor \cite{Akahama1},
the present pressure dependence indicates that $1/T_1$ above 200~K reflects the intrinsic gap, 
whereas $1/T_1$ below 200~K is dominated by relaxation processes associated with the impurity band.

\begin{figure}[t]
\includegraphics[width=0.60\linewidth]{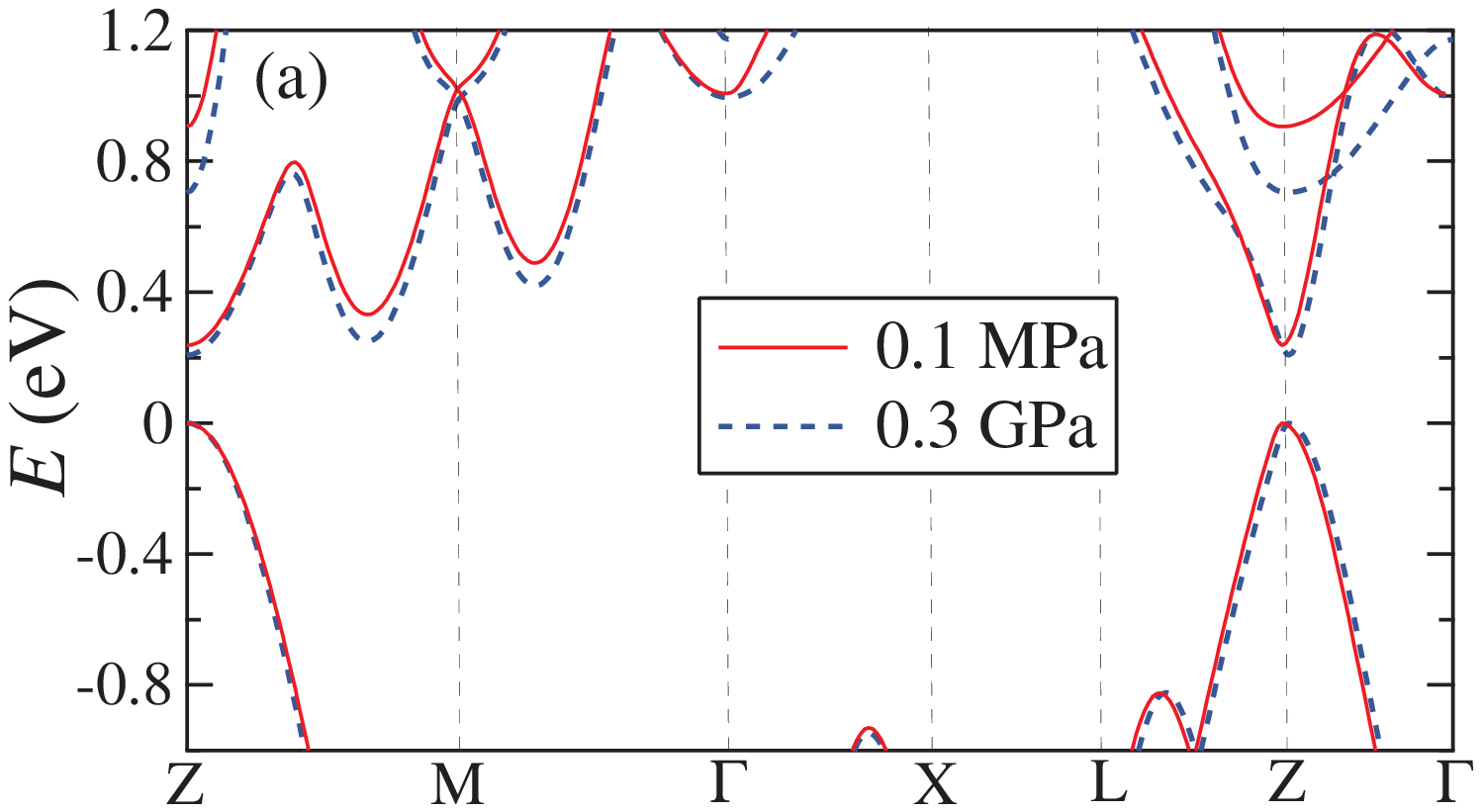}
\includegraphics[width=0.60\linewidth]{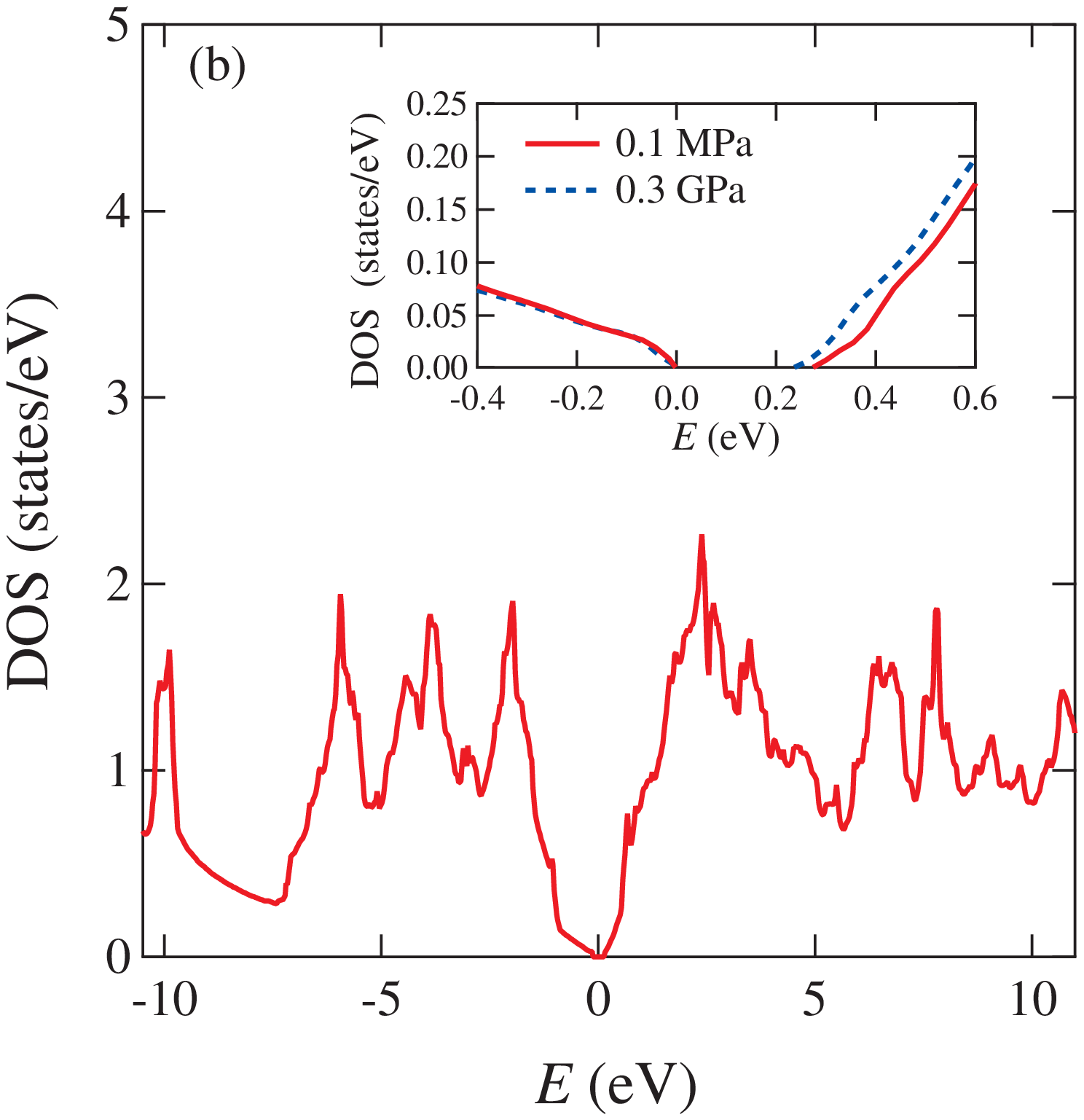}
\caption{\label{fig:epsart}
(color on line)
(a) Calculated band structure near the Fermi energy of BP.
The broken line corresponds to the band dispersion at a pressure of 0.3~GPa.
(b) Energy dependence of total $D(E)$.
The inset shows an expanded view near the Fermi energy.
The broken line corresponds to $D(E)$ at 0.3~GPa.
}
\end{figure}

For nonmagnetic materials, $1/T_1$ is expressed as follows;
\begin{equation}
\label{eq1}
\frac{1}{T_{1}} = \frac{\gamma^2A_{hf}^2}{2}\int_{-\infty}^{\infty} D^2(E) f(\epsilon) (1 - f(\epsilon)) dE,
\end{equation}
where $\epsilon = E - \mu$, $\mu$ is the chemical potential, $A_{hf}$ is the hyperfine coupling constant, $D(E)$ is the density of states, $f(\epsilon)$ is the Fermi-Dirac distribution function, and $\gamma$ is the nuclear gyromagnetic ratio.
One can reproduce the temperature dependence of $1/T_1$
using this equation.
In order to estimate proper $D(E)$, we have performed the band structure calculation with the crystal structure parameters given in Table I.
The obtained band structure near the Fermi energy
is shown by 
the solid line for ambient pressure in Fig. 3(a), which indicates the existence of energy gap of 0.27~eV at the Z point.
The results are in good agreement with previous experimental \cite{Han_ARPES} and theoretical reports \cite{Gong}.

\begin{table}[h]
\caption{
Lattice parameters and positional parameters adopted for the present band structure calculations at ambient pressure and 0.3~GPa \cite{Cartz,Kikegawa}.
}
\begin{center}
\includegraphics[clip,scale=1.2]{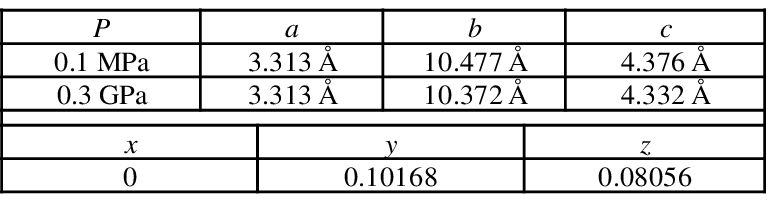}
\end{center}
\end{table}

Figure 3(b) shows the calculated energy dependence of total $D(E)$.
For the actual calculation of $1/T_1$ up to 400~K following Eq.~(1), we may reduce the energy range for the integration to between $\pm 0.5$~eV with respect to the middle of semiconducting gap, i.e. $-0.4 < E < 0.6$~eV.
An expanded view of $D(E)$ for this energy region is presented in the inset of Fig.~3(b).
Moreover we use a simplified model where $D(E)$ in the valence and conduction band region is approximated by linear relations as illustrated in Fig.~4.
In this $D(E)$ model, a Lorentzian-type acceptor band
at the position denoted by $E_{\rm imp}$ is added to the intrinsic bands in order to reproduce the contribution of impurities to the temperature dependence of $1/T_1$ below 200~K as mentioned above.
$\mu$ is determined from the constraint that the total number of electrons $N$ is invariant.
Here, $N$ is defined as follows;
\begin{equation}
N=\int_{-\infty}^{\infty}D(E)f(\epsilon) dE.
\end{equation}

The $1/T_1$ data at ambient pressure were fitted simultaneously using Eqs.~(1) and (2) with $E_g$, $A_{hf}$, and the energy level, amplitude, and width of  impurity band as fitting parameters.
The fitting result is shown by a solid line in Fig.~2.
Also Fig.~4 is depicted based on the resulting fitting parameters.
Here, $E_g = 0.24$~eV is close to the value obtained from the band structure calculation (0.27~eV). Moreover estimated excitation energy between the valence and acceptor bands, $E_{\rm imp} = 25$~meV, and effective impurity concentration, $1 \sim 5 \times 10^{16}$ cm$^{-3}$, seem reasonable compared to previous estimation (18~meV and $2 \sim 5 \times 10^{15}$ cm$^{-3}$, respectively) by the Hall concentration measurement \cite{Akahama1}. Thus the results confirm the validity of the present $D(E)$ model.
The temperature dependence of $\mu$ is shown in the inset of Fig.~2.

\begin{figure}[t]
\includegraphics[width=0.6\linewidth]{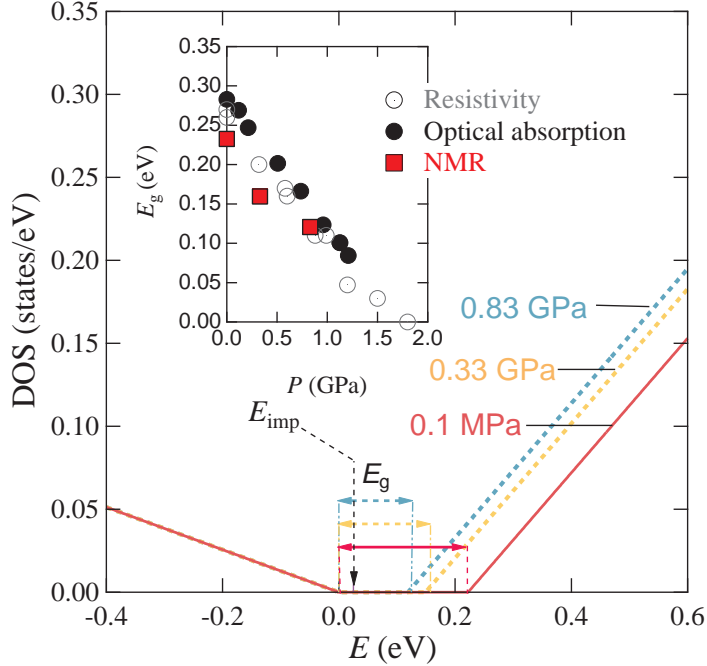}
\caption{\label{fig:epsart}
(color on line)
$D(E)$ model used for the present analyses of $1/T_1$ data.
The valence, conduction, and impurity bands are drawn on the basis of fitting results for $E_g$ and $E_{\rm imp}$.
Inset: pressure dependence of $E_g$ along with previous results by optical absorption (solid circles) and resistivity (open circles) measurements \cite{Akahama2}.
}
\end{figure}

Next, in order to find how $D(E)$ is modified with pressure, we first simulate the effect of applying small pressure on BP by utilizing the band structure calculation.
The previous reports of x-ray diffraction measurements indicate that, as pressure increases, the $b$ and $c$ axes in the orthorhombic structure monotonically shrink, whereas the $a$ axis hardly changes \cite{Cartz,Kikegawa}.
In Table I, the set of lattice parameters, in which the $b$ and $c$ axes are reduced by $\sim 1 \%$, corresponds to the application of a pressure of 0.3~GPa \cite{Kikegawa}.
The calculated effect of pressure on $D(E)$ using these parameters is shown in the inset of Fig.~3(b) (broken lines).
There is no significant pressure induced change in the overall shape of $D(E)$ in the valence and conduction bands, but only $E_g$ is reduced with pressure.
Therefore, in the following analyses of $1/T_1$ at high pressures, we assume that only $E_g$ dependes on pressure.

As shown in Fig.~2, the unique pressure dependences of $1/T_1$, namely the temperature dependence above 200~K becomes steeper with increasing pressure, whereas it is pressure independent below 200~K, are successfully reproduced solely by reducing $E_g$ to 0.16~eV for 0.33~GPa and to 0.12~eV for 0.83~GPa in our $D(E)$ model.
As our $D(E)$ model is constructed on the basis of the result of the band structure calculation, the consistency between the present experimental and theoretical works implies that the band calculations on BP are still reliable in the high pressure region.
This is also supported by the facts that the obtained $E_g$ and its pressure dependence are compatible with those estimated by other experiments as shown in the inset of Fig.~4.

One can gain more detailed information from the analyses using the present $D(E)$ model as described below.
The obtained value of $1/T_1$ is decomposed into components originating from the intrinsic semiconducting band structure and from the impurity band.
We divide Eq.~(1) into the following two equations;
\begin{eqnarray}
\scalebox{0.90}{$\displaystyle
\left( \frac{1}{T_{1}} \right)_{\rm intr}
= \frac{\gamma^2 A_{hf}^2}{2} \left[
\frac{N_{\rm intr}}{N_{\rm intr} + N_{\rm accp}} \int_{-\infty}^{0} D^2(E) f(\epsilon) (1-f(\epsilon)) dE \right.    
\left. + \int_{E_{g}}^{\infty} D^2(E)f(\epsilon)(1-f(\epsilon)) dE \ \right]
$}
\end{eqnarray}
\begin{eqnarray}
\scalebox{0.95}{$\displaystyle
\left( \frac{1}{T_{1}} \right)_{\rm accp} = \frac{N_{\rm accp}}{N_{\rm intr} + N_{\rm accp}} \frac{\gamma^2 A_{hf}^2}{2}\int_{-\infty}^{0}D^2(E)f(\epsilon)(1-f(\epsilon)) dE
$}
\end{eqnarray}

\noindent
where $N_{\rm intr}$ and $N_{\rm accp}$ are the numbers of electrons which are thermally excited from the valence band to the conduction band and to the acceptor band, respectively.
Equation~(3) describes the $T_1$ relaxation associated with the intrinsic semiconducting band structure, and Eq.~(4) describes the $T_1$ relaxation caused by the holes created in the valence band through the valence-acceptor band transition.
Here we take only major relaxation processes into account, and minor contributions, such as $T_1$ components caused by electrons in the acceptor band as well as electrons in the valence band through the acceptor-conduction band transition, are neglected due to the tiny $D(E)$ in the acceptor band.
The decomposed $(1/T_1)$'s are represented by the broken lines in Fig.~2.
$1/T_1$ below 200~K is dominated by $(1/T_1)_{\rm accp}$, which is obviously the main reason for the behavior insensitive to pressure in this temperature region.
In contrast, $(1/T_1)_{\rm intr}$, having steeper temperature dependence than $(1/T_1)_{\rm accp}$, becomes dominant above 300~K.
The results reveals that the suppression of $E_g$ with pressure induces the shift of the $(1/T)_{\rm intr}$ vs $T$ curve toward lower temperatures, so that $1/T_1$ is pushed up in the high temperature region.

The temperature and pressure dependences of $\mu$ have been also evaluated using the $D(E)$ model.
As shown in the inset of Fig.~2, $\mu (T)$ at ambient pressure monotonically increases with temperature, but at the high temperature limit it does not reach 0.1~eV, lower than $E_g /2 \sim 0.12$~eV, reflecting the asymmetric $D(E)$ with respect to the middle of  intrinsic gap.
As pressure increases, $\mu$ above 250~K is suppressed, which is induced by the shrinkage of $E_g$.
Although the present estimation was carried out only within the semiconducting state, the information on $\mu$ will be crucial for examining
whether or not Dirac fermions are formed at higher pressures in BP.

In addition to the semiconducting phase, we have also performed the $T_1$ measurement at 1.63~GPa, seemingly just above $P_c$ compared with the value of $P_c$ reported previously.
The obtained $1/T_1$ is one order of magnitude larger than the data at ambient pressure in the whole temperature range (see Fig.~2).
Since $(1/T_1)_{\rm accp}$ is independent of pressure as explained above, the enhancement of $1/T_1$ is unambiguously ascribed to a change in the intrinsic band structure, namely the density of conduction electrons is considerably increased due to the collapse of  semiconducting gap.
This is consistent with the evolution of semimetallic properties evidenced by the decrease in the resistivity \cite{Akahama2,Xiang,Akiba} and the observation of Shubmikov-de Haas oscillation \cite{Xiang,Akiba} above about 1~GPa.
The result suggests that the temperature dependence of $1/T_1$ at 1.63~GPa is approximately proportional to $T^3$, which is expected when the Dirac corns are formed.
Indeed similar behavior is observed in the Dirac compound $\alpha$-(BEDT-TTF)$_2$I$_3$ \cite{Hirata} and in the Weyl semimetal TaP \cite{Yasuoka}.
However the present $D(E)$ model does not successfully reproduce the whole temperature dependence of $1/T_1$ at 1.63~GPa, suggesting the importance of more careful estimation of $D(E)$ and $\mu$.
We expect that further systematic NMR measurements as a function of pressure near $P_c$ will be useful to improve the $T_1$ simulation to extract detailed information about the formation of Dirac fermion in BP.

In summary, we have for the first time carried out $^{31}$P-NMR measurements of BP at ambient pressure and high pressures up to 1.63~GPa seemingly exceeding $P_{c}$.
The obtained temperature and pressure dependences of $1/T_1$ are well reproduced by the calculation based on the $D(E)$ model derived from the band structure calculations, giving a good confirmation that the band calculations on BP are fairly reliable.
The results enable us to know how $D(E)$ near the Fermi level changes with increasing pressure, and the obtained pressure dependence of $E_g$ is in good agreement with previous reports.
Moreover the successful analyses in this study provide more detailed information in the semiconducting state, including the decomposed intrinsic and extrinsic contribution to the $T_1$ relaxation and the pressure and temperature dependences of $\mu$, which will be useful to examine the realization of Dirac fermions under pressure in BP.

\begin{acknowledgments}
We are grateful to Profs. T.~Mutou and Y.~Hasegawa for valuable discussions and Prof. T.~Nomura for help with the band calculation.
This work was supported by JSPS KAKENHI (Grant No. 18H04331).
\end{acknowledgments}



\end{document}